\documentclass{edp-conf}
\usepackage{graphicx}
\begin{document}

\TitreGlobal{SF2A 2005}

\title{ALMA and Cosmology}
\author{Combes, F.}\address{Observatoire de Paris, LERMA, 61 Av. de l'Observatoire, F-75014, Paris}
\runningtitle{ALMA and Cosmology}
\setcounter{page}{237}
\index{Combes, F.}
%
\maketitle
\begin{abstract}
ALMA is a privileged instrument to tackle high redshift galaxies, 
due to the negative K-correction in the millimeter domain. 
Many dusty star-forming galaxies, invisible in the optical or NIR, 
will be detected easily through the peak of their emission in the 
FIR redshifted in the submm between z=10 and z=5. Their mass and dynamics 
will be determined through the CO lines, together 
with the efficiency of star formation. Normal intervening galaxies at all 
z will be studied through absorption lines in front of quasars, 
exploring the dense tail of the column density spectrum. CMB anisotropies 
could be detected at the arcsecond scale, the secondary 
effects (SZ, Vishniak-Ostriker) could test the re-ionization and the nature 
of dark energy. The detection of the SZ effect on a few 
arcsec scales will allow to map in detail clusters and proto-clusters.
\end{abstract}
%
\section{High redshift galaxies}

Star forming galaxies emit most of their light in the far infrared around 100 microns,
due to emission from dust heated by the recent stars formed (e.g. Sanders \& Mirabel 1996).
At high redshift, the peak of the SED (spectral energy distribution) of galaxies
is moving to the submillimeter range, which is at the origin of the negative K-correction
effect: it is as easy to detect the continuum emission from $z=1$ and $z=10$ objects.
  Already at the present sensitivity of submm instruments (SCUBA on JCMT,
MAMBO on IRAM-30m), a few hundreds of galaxies have been detected at high
redshift, and many of them are not identified in the visible domain. Above a
flux of 1mJy, deep surveys detect about 1 source per arcminute (e.g. Greve et al 2004).
 These SMG (submillimeter galaxies) are  very actively forming
stars,  as luminous as ULIRGs.
 
 At the moment, it is not possible to detect more modest starbursts,
and in particular the typical Lyman break galaxies at redshift around 3 
are not detected. With the 100 times better sensitivity of ALMA,
they will all be detected, and the number of sources in deep surveys
is expected to be multiplied by 2 orders of magnitude (100 objects per
arcminute).  For very high redshifts objects (z $>$ 7), the sensitivity of
ALMA will be larger than in the visible/NIR, larger than for the JWST.

\section{Star formation history}

The star formation history is quite uncertain at high z
(e.g. Genzel \& Cesarsky 2000), due precisely to the ill-known
amount of extinction suffered by high-z galaxies. The
indication given by the SMG pushes towards a high contribution,
corresponding to a large fraction of the cosmic infrared background
accounted by starbursting enshrouded galaxies (Elbaz et al. 2002).
In the local universe, the star formation rate is dominated
by normal (quiescent) galaxies. It begins to be dominated
by LIRGs (with L $>$ 10$^{11}$ L$_\odot$) at $z=0.7$,
and it is likely that ULIRGs (L $>$ 10$^{12}$ L$_\odot$)
dominate at very large redshift (z$>$ 2.5).
The SMGs detected until now have a median redshift of
$z=2.2$, and the dust-corrected UV luminosities can underestimate
their bolometric luminosity by 2 orders of magnitude
(Chapman et al 2005).

In addition to dust continuum emission, ALMA will detect the 
CO lines, which are much more difficult to detect,
due to the absence of negative K-correction 
 (e.g. Combes et al 1999).
  With the present sensitivity, it is only possible to
detect high-z huge starbursts, and their star formation 
efficiency (SFE) is quite surprising (Greve et al 2005).
 Figure 1 (left) shows that their SFE, as traced from the ratio
of the infrared luminosity (heated dust emission) to
the amount of molecular gas present (CO luminosity), is higher 
than for LIRGs and ULIRGs at lower redshift. 
Since ULIRGs are explained by major mergers, 
SMG could be mergers of high-z galaxies with
a lower bulge-to-disk mass ratio,  for which
a merger is more violent in triggering disk instability
(e.g. Mihos \& Hernquist 1996).
From a compilation of all SMGs detected up to now, a typical SFR
of 700 M$_\odot$/yr, and a starburst phase of 40-200 Myr can be deduced
for SMGs (Greve et al 2005). Their typical masses are 0.6 M*, and their number is 
compatible with hierarchical galaxy formation scenarios (Baugh et al 2005).

\section{Dark matter and intervening absorptions}

The CO lines at high redshift will allow to determine 
more reliably than in the visible domain the kinematics of galaxies.
They will give information on the dark matter
distribution in galaxies, in the absence of the HI line, 
(to be obtained later with SKA). The CO Tully-Fisher relation 
is even more accurate, since the CO line width is less  broadened
by galaxy interactions than the HI width (Lavezzi \& Dickey 1998,
Tutui \& Sofue 1997).

Absorption lines are much more sensitive than emission lines
to cold gas along the line of sight, provided that a strong 
continuum source is detected. With ALMA, the number of background
sources available in the millimeter domain will be multiplied
by 2 orders of magnitude, and it will be possible to explore
the chemistry as a function of redshift, the CMB temperature
versus z, or to check the variations of the fine structure constant,
or the ratio of electron to proton mass, etc.. (e.g. Wiklind \& Combes
1998).

\begin{figure}
   \begin{minipage}{6cm}
      \centering \includegraphics[width=5.5cm]{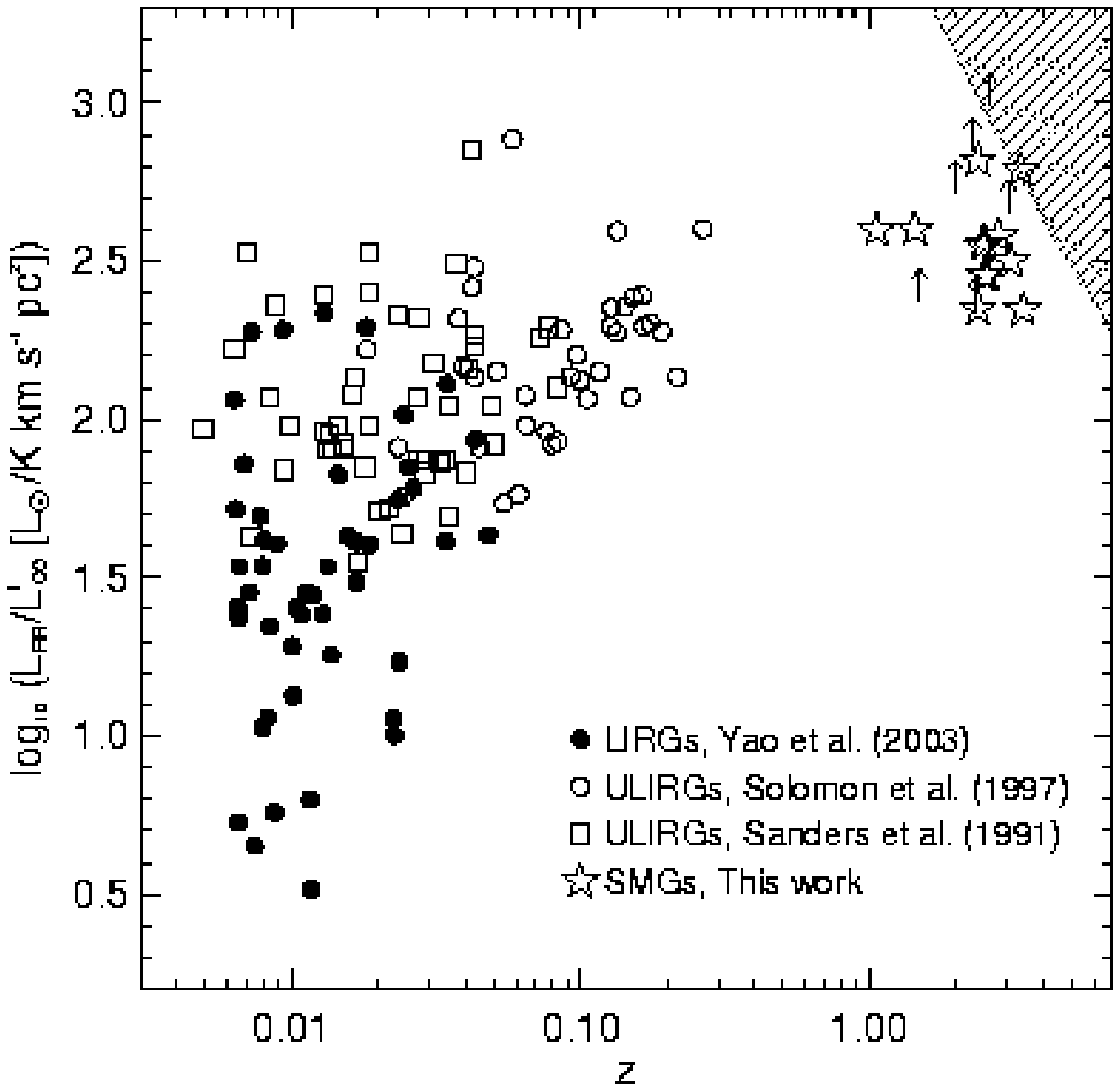}
   \end{minipage}
\hfill
   \begin{minipage}{6cm}
      \centering \includegraphics[width=5.5cm]{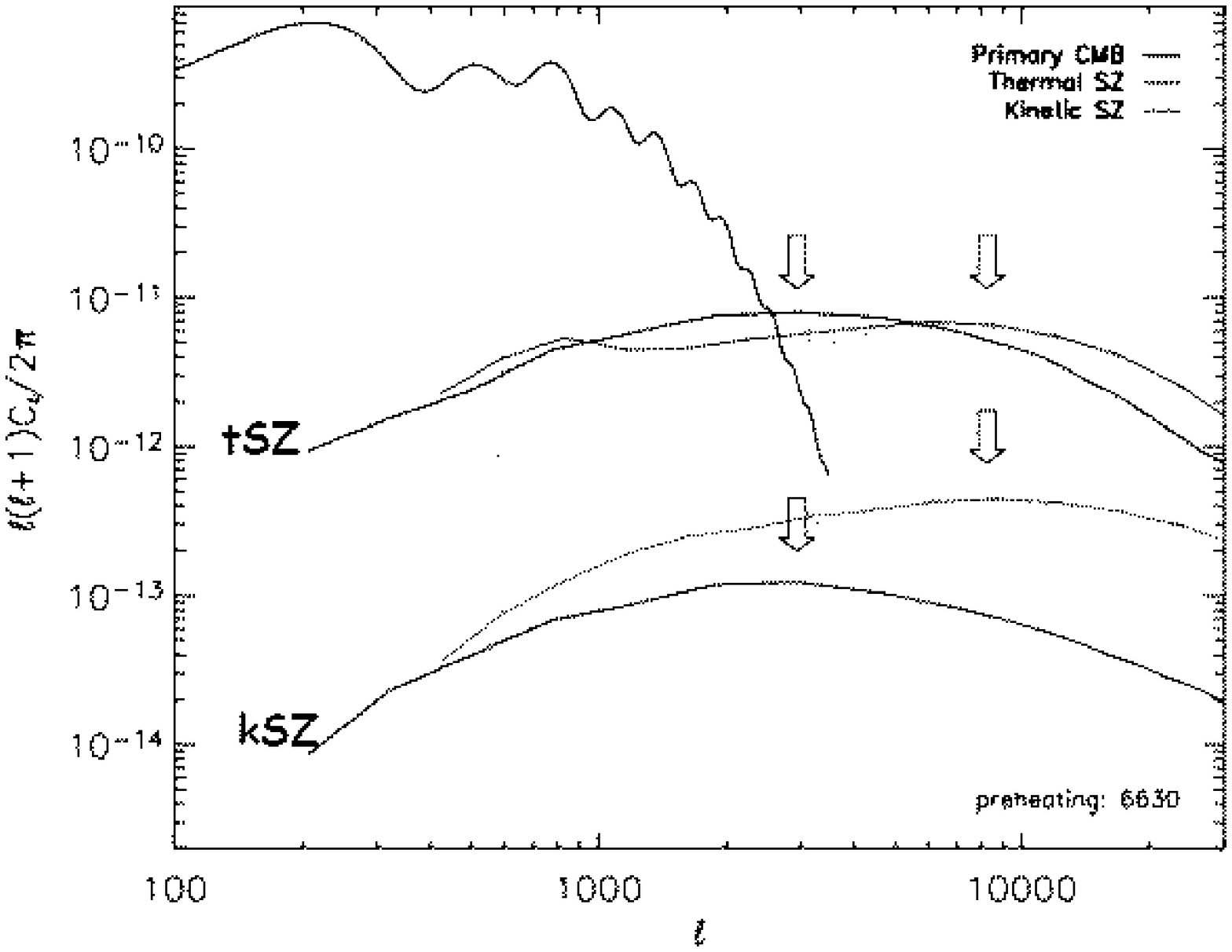}
   \end{minipage}
      \caption{{\it \bf Left}: 
 The star-formation efficiency, traced by L$_{FIR}$/L$_{CO}$,
as a function of redshift. The SMG (SubMillimeter Galaxies) are plotted as
stars, and compared to LIRGs (full symbols) and ULIRGs (open symbols). 
From Greve et al. (2005). 
$\quad \quad \quad \quad \quad \quad$
   {\it \bf Right}: 
      Predictions from numerical simulations of clusters
of the thermal SZ effect (noted tSZ), and the kinetic SZ effect (noted kSZ),
in the diagram of anisotropy power as a function of scale. The pattern
of CMB primary anisotropy is plotted for comparison.
From da Silva et al (2004).  }
   \end{figure}

\section{SZ effect and galaxy clusters at all z}

Since its first detection with the Ryle interferometer 
(Jones et al 1993), the thermal SZ has been detected in
 about 100 clusters (e.g. reviews by Birkinshaw 1999, Carlstrom et al 2002),
with interferometers (BIMA, OVRO) or single dishes (Diabolo
on IRAM-30m, Scuba on JCMT, Bolocam on CSO, ACBAR on Viper).
The SZE is a precious tool for cosmology, since the detection rate is
almost independent of redshift, it depends only on the mass of the cluster.
With more sensitive instruments, the mass limit will decrease, and 
angular resolution is then needed, to map clusters at high redshift.
ALMA will detect the kinetic SZ (undetected up to now), 
which is expected 2 orders of magnitude
lower (cf Figure 1, right), and trace possible cluster rotation.

There are two kinds of scientific goals that could be
pursued with the SZE:
\begin{itemize}
\item
Determine the cosmological parameters: by obtaining a
cluster-based Hubble diagram, constrain w (the state equation
of dark energy), $\Omega_m$
and $\sigma_8$ through 
cluster counts as a function of redshift,
and obtaining the variation of 
T$_{cmb}$ with (1+z).
\item
Study the physics of clusters, by mapping the hot gas
across clusters, and their radial velocity (through the kinetic SZ),
obtaining the baryon fraction, the mass-to-temperature relation, etc..
\end{itemize}

In 2010, at the start of ALMA, 5000 clusters 
will have been observed with Planck, but essentially at z$\sim$ 0.
A few hundreds will be detected at higher z with
the near future experiments like AMIBA, SZA, APEX etc.
and several dozens with spectral measurements.
ALMA will be able to make detailed maps, with
resolution better than a few arcsecs, with 
much increased sensitivity, reaching the cluster physics
(shocks, cooling flows, cold fronts, map of Te).
It will be hard to make blind surveys however, 
since ALMA has too small field of view. The future
X-ray satellite Constellation-X/Xeus will be needed to 
identify the clusters.
ALMA will be able to detect polarisation signals, 
which are quite weak, and require
high angular resolution.

\section{Other secondary anisotropies}

The Ostriker-Vishniac effect can provide a unique probe 
of the epoch of reionization, and on galaxy formation.
Large ionized clouds at redshift z between 7 and 10
will produce anisotropies at the arcsec scale
(Ostriker \& Vishniac 1986).
Estimations of the effect have been computed,
with ionized gas sizes of
D=2.5-30kpc, i.e. angles between 1 and 6 arcsec at $z=9-30$
(Peebles \& Juszkiewicz 1998, Jaffe \& Kamionkowski 1998a).
The effect is dominated by the kinetic term
(V $\sim$ 600km/s), and the expected signal is of the order
of 30 micro K, detectable with ALMA.


\end{document}